%% file: main.tex
\documentclass[journal,twoside,web]{ieeecolor}
\usepackage{generic}
\usepackage{cite}
\usepackage{amsmath,amssymb,amsfonts}
\usepackage{algorithmic}
\usepackage{graphicx}
\usepackage{subfigure}
\usepackage{algorithm,algorithmic}
\usepackage{hyperref}
\hypersetup{hidelinks=true}
\usepackage{textcomp}
\usepackage{booktabs}
\usepackage{xcolor} 
\usepackage[margin=0.63in]{geometry}
\usepackage{titlesec}
\titlespacing*{\section}{0pt}{0.25ex}{0.0ex}
\titlespacing*{\subsection}{0pt}{0.25ex}{0.25ex}
\titlespacing*{\subsubsection}{0pt}{0.25ex}{0ex}

\def\BibTeX{{\rm B\kern-.05em{\sc i\kern-.025em b}\kern-.08em
    T\kern-.1667em\lower.7ex\hbox{E}\kern-.125emX}}
\begin{document}
\title{\huge Advancing Problem-Based Learning in Biomedical Engineering in the Era of Generative AI}
\author{Micky C. Nnamdi, J. Ben Tamo, Benoit Marteau, Wenqi Shi$^{*}$, May D. Wang$^{*}$
\thanks{This research has been supported by funding of Wallace H. Coulter Distinguished Faculty, Petit Institute Faculty Fellow, Inclusive STEM Teaching Fellow,  Leaders for Teaching and Learning Fellow, Amazon, and Microsoft Research to Professor May D. Wang. 
M.C.N., J.B.T., and B.M. are with the Department of Electrical and Computer Engineering, Georgia Institute of Technology, Atlanta, GA, USA;
W.S. is with the Department of Health Data Science and Biostatistics
at Peter O'Donnell Jr. School of Public Health, UT Southwestern Medical Center, Dallas, TX, USA;
M.D.W. is with the Wallace H. Coulter Department of Biomedical Engineering, Electrical and Computer Engineering, Georgia Institute of Technology and Emory University, Atlanta, GA, USA. 
* two co-corresponding authors e-mails: wenqi.shi@utsouthwestern.edu; maywang@gatech.edu.}}

\maketitle

\begin{abstract}
\textbf{\textit{Contribution:}} This paper presents a modularized, implementation-ready Problem-Based Learning (PBL) framework tailored to biomedical artificial intelligence (AI), with safe use of Generative AI (GenAI) for knowledge summarization and coding support. A replication package (syllabi, milestones, rubrics, team procedures, and AI-usage templates) accompanies the framework.

\textbf{\textit{Background:}} PBL has significantly impacted Biomedical Engineering (BME) education since its introduction in the early 2000s, effectively enhancing student learning through critical thinking and real-world problem-solving. However, traditional PBL approaches demand considerable faculty resources, domain-specific expertise, and continuous curricular updates to align with rapidly evolving biomedical technologies. Recent advancements, including the 2024 Nobel Prizes awarded for AI-enabled discoveries, highlight the importance of comprehensive student training in AI. As biomedical engineering rapidly converges with AI, integrating effective AI education into established curricula is necessary but faces challenges. These include diverse student backgrounds, limited personalized mentoring, constrained computational resources, and difficulties in safely scaling hands-on experiments due to privacy and ethical concerns associated with biomedical data.

\textbf{\textit{Intended Outcomes:}} This study aims to improve students' grasp of biomedical AI concepts, increase proactive learning and teamwork, and address equity and scalability concerns in AI education through authentic real-world problem-solving experiences.

\textbf{\textit{Application Design:}} A three-year case study (2021–2023) across the Georgia Institute of Technology and Emory University engaged 248 students in interdisciplinary teams to address real biomedical AI problems. The curriculum positioned GenAI as both a class topic and a learning tool, governed by strict disclosure, source-anchoring, verification, and version-logging policies.

\textbf{\textit{Findings:}} This implementation coincided with measurable improvements in learning outcomes, evidenced by high problem-solving productivity (16 student-authored peer-reviewed publications), positive peer evaluations, and the successful development of innovative computational methods for real biomedical problems. This strategy not only prepares students for future healthcare innovation but also systematically addresses educational disparities and resource limitations inherent to conventional learning approaches. The study presents a practical and scalable roadmap for biomedical engineering departments aiming to integrate robust AI education into their curricula.
\end{abstract}

\begin{IEEEkeywords}
Biomedical Education,  generative artificial intelligence, large language models, problem-based learning
\end{IEEEkeywords}

\input{sec/1-intro}

\input{sec/2-relatedworks}
\input{sec/3-casestudy}
\input{sec/4-genAI}
\input{sec/6-conclusion}

\section*{Conflict of Interest}
All authors declare no conflicts of interest.

\section*{Author Contribution}
M.C.N., J.B.T., B.M., and W.S. contributed to the study design, data collection, statistical analysis, result evaluation, and writing of the manuscript, including figures and tables. M.D.W. contributed to the course content development, study design, result evaluation, and writing of the manuscript. All authors reviewed the manuscript. 

\section*{Acknowledgment}
We truly appreciate Georgia Tech Center for the Enhancement of Teaching and Learning (CETL) for inviting us to share our novel design of Problem-Based Learning (PBL) for Technical Specialty Courses in Celebrating Teaching Day from 2019 to 2025, and for large GT community to provide encouragement and feedback. We truly appreciate multiple Faculty Fellow supports to Prof. May D Wang from CETL and the Wallace H Coulter Department of Biomedical Engineering. To implement this pedagogical innovation, we truly appreciate the contributions from many PhD students who specialized in biomedical health informatics: Dr. Ryan Hoffman, Dr. Li Tong, Dr. Hang Wu, Dr. Felipe Giuste, Dr. Yuanda Zhu, and Yishan Zhong. We thank all the students and teaching assistants from Georgia Tech and Emory University who participated in this study.

\bibliographystyle{IEEEtran}
\bibliography{IEEEabrv,ref}

\end{document}

%% file: sec/1-intro.tex
\section{Introduction}
\label{sec:introduction}

\IEEEPARstart{B}{iomedical} education and research have been advancing together, each reinforcing the other’s development \cite{patel2009cognitive, irby2010calls}. Problem-Based Learning (PBL) represents a prime example of such integration in Biomedical Engineering (BME) training \cite{savery2015overview, savery1995problem, patel2009cognitive}. Unlike traditional lecture-based curricula, PBL encourages interdisciplinary integration and practical problem-solving. It centers learning around authentic, open-ended, real-world biomedical problems, where small teams drive self-directed inquiry and collaborate to develop iterative solution design. The Wallace H. Coulter Department of Biomedical Engineering at Georgia Institute of Technology (Georgia Tech) and Emory University pioneered PBL education in BME and has received the National Academy of Engineering’s Gordon Prize for Innovation in Engineering and Technology Education in 2019 \cite{gordonprize2019}. 

Students trained through PBL demonstrate stronger clinical reasoning skills compared to their counterparts in conventional curricula, highlighting the effectiveness of PBL in BME education \cite{hmelo2004problem}. Newstetter et al. showed that PBL significantly improves the translation of theoretical concepts into practical solutions, an essential skill given the biomedical field’s emphasis on bench-to-bedside translation  \cite{newstetter2010design}. Additionally, comprehensive studies by Linsenmeier and Saterbak documented that a robust PBL methodology increased student retention and satisfaction in BME programs \cite{linsenmeier2020fifty}. The alignment between PBL and BME arises naturally from the field’s interdisciplinary requirements: through engagement with real-world scenarios, students integrate complex multi-disciplinary engineering principles, life sciences, clinical medicine, and computational skills in ways that mirror professional practice \cite{barrows2000problem, beddoes2010identifying}. However, traditional PBL approaches demand considerable faculty resources, domain-specific expertise, and continuous curricular updates to align with rapidly evolving biomedical technologies.
\begin{figure*}[ht]
    \centering
    \includegraphics[width=0.93\linewidth]{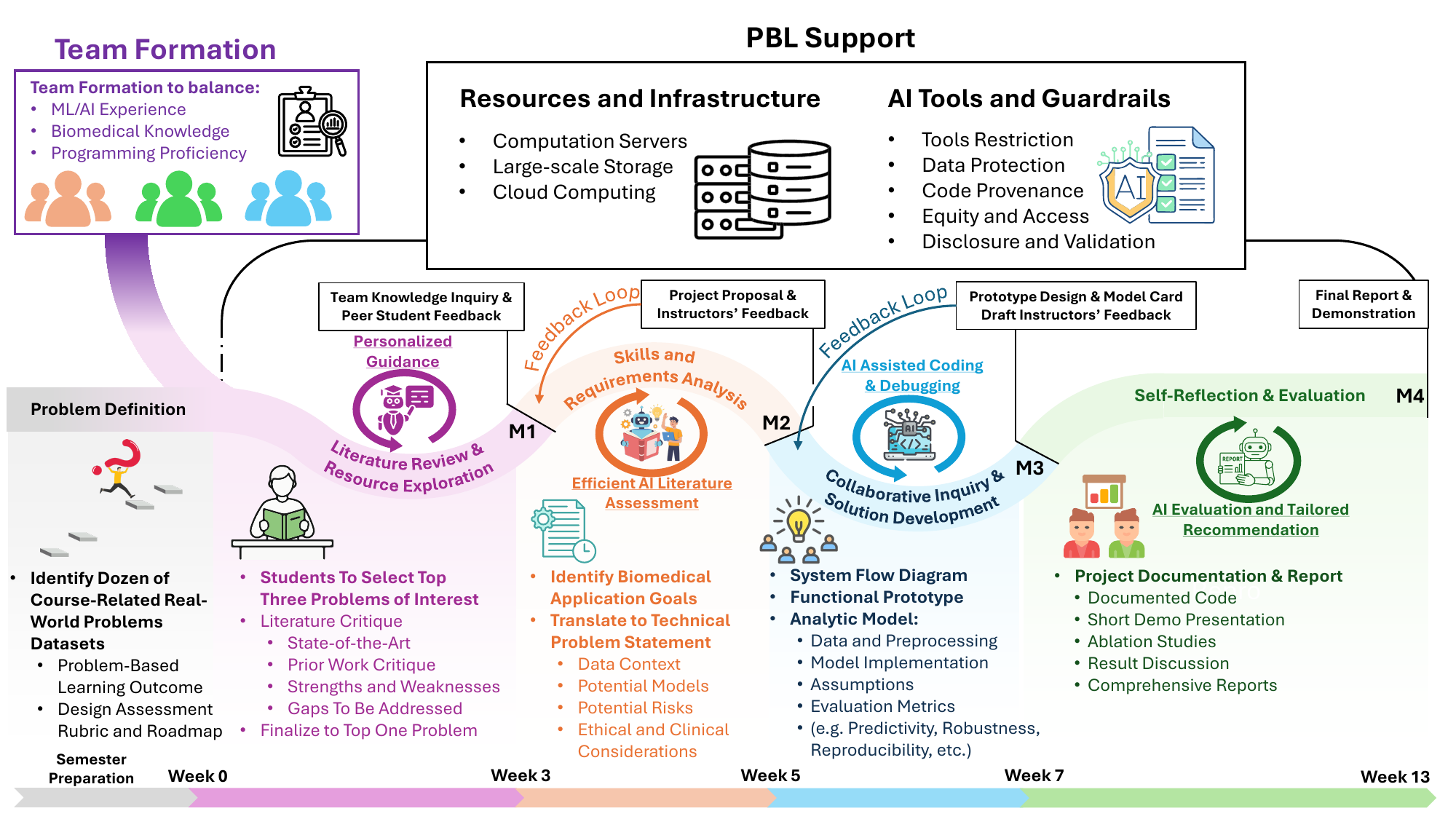}
    \caption{Modularized PBL+AI framework for biomedical engineering education. The pipeline consists of key stages including Problem Formation, where authentic biomedical briefs and curated datasets anchor student inquiry; Knowledge Inquiry Supported by AI, where GenAI scaffolds literature synthesis, coding initialization, and reflection without replacing reasoning; Problem-solving, where teams design experiments, refine models, and evaluate reproducibility and robustness; and Presentation, where learning outcomes are consolidated through written reports, oral presentations, and peer review}
\label{fig:pbl2}
\end{figure*}
Thus, the recent advances in Artificial Intelligence (AI), particularly Generative AI (GenAI) such as Large Language Models (LLMs) capable of producing text and code from prompts, have introduced new possibilities for PBL in technical specialty courses. These GenAI systems can synthesize and summarize existing knowledge, provide students with rapid access to state-of-the-art references~\cite{lu2024large, mishra2024use, he2025survey}. With proper disclosure, instructor oversight, and ethics training, it can support literature navigation, formative reflection, and code refactoring, without replacing human reasoning. GenAI helps students move more quickly from information-gathering to higher-order activities such as critical evaluation, synthesis, and innovation. Importantly, AI in PBL is not to provide direct answers that bypass the learning process, but to assemble a reliable baseline of current knowledge. Thus, it strengthens rather than weakens the pedagogical intent of PBL.

In the current era of AI, BME programs provide basic AI exposure; however, student understanding often lacks technical depth. Conversely, students in computer science-related programs may possess deep knowledge of AI but lack meaningful biomedical context. This disconnect limits the applicability of AI to real-world biomedical problems~\cite{sharma2024integrating, hudson1999neural, naskar2024biomedical}. Thus, current curricula do not consistently equip students with AI competence in healthcare and BME. This study addresses this gap by presenting a modularized, operational framework that integrates GenAI into PBL for biomedical AI education. A three-year case series (2021–2023) across Georgia Tech and Emory University engaged 248 undergraduate and graduate learners. The curriculum featured multiple modules designed to develop student AI proficiency alongside biomedical competencies (ethics, regulatory awareness, clinical integration).
This study makes three major contributions: (i) it created an implementation-ready architecture that positions GenAI as a guarded knowledge-summarization and coding-support module within the PBL cycle; (ii) it produced a replication package consisting of syllabi, milestones, rubrics, team-formation procedures, and prompt/disclosure templates; and (iii) it provided empirical evidence from historical cohort outcomes and exploratory analyses, supplemented by peer-assessment indicators, to calibrate claims about learning. The instructional workflow treated GenAI as both a class topic and a learning tool, with risk-mitigation strategies (disclosure, source anchoring, verification, and version logging) embedded in the process.

%% file: sec/2-relatedworks.tex
\begin{table*}
\centering
\caption{Examples of Problem-Based Learning in Biomedical Engineering Education}
\label{tab:pbl-implementations}
\resizebox{\linewidth}{!}{
\begin{tabular}{l|l|l}
\hline
\textbf{Institution} & \textbf{Program} & \textbf{Key Features} \\
\hline
Georgia Tech \& Emory University & BME Curriculum & Integrated design challenges, Full BME curriculum \\
Stanford University & Biodesign Program & Clinical immersion, Needs identification \\
Johns Hopkins University & BME Undergraduate Design Team & 18-months design team projects, Clinical sponsor\\
Duke University & BME Design Fellows & Open-ended problems, Industrial partnerships  \\
University of Michigan & Clinical Immersion & Early clinical and translational science exposure, Healthcare system focus  \\
Harvard-MIT & HST Medical Engineering and Medical Physics (MEMP) & Cross-disciplinary problems, Patient-centered  \\
UC San Diego & DesignLab & Global health challenges, Humanitarian focus \\
Purdue University & EPICS in BME & Service learning integration, Community partners\\
\hline
\end{tabular}
}
\end{table*}

\begin{table*}[!ht]
\centering
\caption{Comparative Overview of Biomedical AI Education Programs}
\label{tab:ai-education}
\resizebox{\textwidth}{!}{
\renewcommand{\arraystretch}{1.5}
\begin{tabular}{p{4cm}|p{4.5cm}|p{6cm}|p{3cm}}
\hline
\textbf{Institution / Program} & \textbf{Educational Approach} & \textbf{Key AI Focus Areas} & \textbf{Integration Type}\\
\hline
Stanford University – AI in Healthcare & Specialized certificate program & Clinical safety, medical applications of AI, clinical decision-making support & Specialization\\
\hline
Harvard T.H. Chan School of Public Health – Responsible AI for Healthcare & Executive education certificate program & Ethical AI frameworks, responsible implementation, regulatory considerations & Specialization\\
\hline
Johns Hopkins University – AI in Medicine & Master's degree specialization & Computational medicine, biomedical imaging analysis, machine learning for clinical decision support & Integrated specialization\\
\hline
Duke University – AI Health & Multidisciplinary certificate program & Population health analytics, precision medicine, interdisciplinary AI methods & Hybrid\\
\hline
UCSF AI4ALL – Summer Immersion Program & Experiential summer institute & Applied biomedical AI projects, inclusive AI education, hands-on interdisciplinary learning & Hybrid (Experiential)\\
\hline
Carnegie Mellon University – AI Scholars Program & Pre-college immersive summer program & AI fundamentals, machine learning, robotics, computational thinking & Standalone\\
\hline
Imperial College – AI for Healthcare & Interdisciplinary collaboration and academic-industry partnerships & Medical imaging, diagnostic AI, clinical decision support systems & Integration\\
\hline
University of Toronto – Temerty Centre for AI Research and Education in Medicine (T-CAIREM) & Integrated clinical and research initiative & Predictive analytics, machine learning in clinical decision-making, translational AI research & Integration\\
\hline
ETH Zurich – Biomedical AI Workshop Series & Seminars and workshop-based education & AI ethics, healthcare AI regulation, validation of biomedical AI models & Workshop Series\\
\hline
\end{tabular}}
\end{table*}

\section{Related Work}
The intersection of AI and biomedical education has been growing over the past decade. This section reviews two central topics: (i) the development of biomedical AI education programs and (ii) the emerging role of GenAI within biomedical education.

\subsection{Biomedical AI Education}
Educational programs in biomedical AI have experienced considerable changes due to the widespread adoption of AI technologies across healthcare. Historically, traditional biomedical curricula generally emphasized foundational biological and clinical knowledge learning, with computation treated as secondary \cite{maleki2024role, khalifa2024artificial}. However, the accelerated growth and broader integration of AI into healthcare have necessitated significant curriculum adaptations. Medical schools and biomedical engineering departments started creating specialized curricula that systematically integrate clinical knowledge with core AI competencies (Tables~\ref{tab:pbl-implementations} and \ref{tab:ai-education}). For example, Stanford Medical School developed a specialized biomedical AI track explicitly designed to train students with dual expertise in clinical medicine and Machine Learning (ML) methodologies \cite{ngo2022cases, altman2007biomedical, iyer2024advancing}. Such initiatives illustrate the transition toward more cohesive and rigorous approaches in biomedical AI education. Current educational practices in biomedical AI can be grouped into three distinct frameworks:

\begin{itemize}
\item \textbf{Integration Model:} AI concepts and computational tools are systematically integrated into existing biomedical curricula. This approach introduces AI within the context of relevant clinical and biomedical scenarios, ensuring students understand AI applications directly in clinical practice~\cite{sharma2024integrating}.
\item \textbf{Specialization Model:} Dedicated academic tracks or specialized concentrations focus explicitly on biomedical applications of AI. Typically, this model demands additional, intensive coursework in computer science, machine learning, data analytics, and their biomedical applications, thus preparing students for specialized careers at the intersection of medicine and AI \cite{jimenez2024problem}.
\item \textbf{Hybrid Model:} Combining elements of integration and specialization approaches, hybrid models offer core biomedical curricula alongside optional advanced AI coursework and research opportunities. This model provides flexibility to accommodate diverse student backgrounds and interests, enabling tailored education pathways \cite{lopez2023problem}. 
\end{itemize}

Within this framework, the proposed PBL with GenAI approach aligns with the Hybrid Model, as it uses authentic biomedical problem briefs to organize AI learning. In contrast to the Integration Model, this design introduces multiple specialized concepts (e.g., model cards, ethics statements) and analytics-ready assessment rubrics to make AI practice observable and ratable. Compared with specialization tracks, the approach prioritizes domain-grounded inquiry and cross-disciplinary teaming over stand-alone algorithm courses. The Methods section details the participants, instruments, and reliability measures used to support replication and comparative evaluation.

\subsection{Biomedical Education in the GenAI Era}
GenAI, particularly LLMs, introduces new possibilities relevant to biomedical education. Reported uses include: (i) \emph{on-demand tailored tutoring and explanation} (structured concept review, code comments); (ii) \emph{literature critique and knowledge inquiry} (querying and summarizing primary sources); (iii) \emph{data workflows and coding packages} (package understanding, refactoring, unit testing, exploratory analysis); (iv) \emph{problem-solving and implementation support} (scenario generation and feedback for case discussions); and (v) \emph{formative assessment} (feedback aligned to rubrics) \cite{abd2023large, lu2024large, mishra2024use, he2025survey, li2023llava, amiri2025project}. When paired with disclosure requirements and instructor guardrails, such tools can support PBL stages (ideation, planning, reflection) without supplanting student reasoning.

However, recognizing the challenges and limitations of GenAI is essential. LLMs may produce plausible but incorrect content (hallucinations), reflect training data biases, or obscure provenance, raising validity and equity concerns for assessment and access \cite{parente2024generative}. Responsible use in coursework therefore requires (i) explicit AI use policies and disclosure logs, (ii) source-anchored verification (e.g., citation checks, retrieval of primary evidence), (iii) alignment of assignments with \emph{process} artifacts (planning notes, error analyses) rather than final answers alone, and (iv) rater training with reliability checks to prevent uncritical acceptance of AI-mediated work.

%% file: sec/3-casestudy.tex
\section{Materials and Methods}

\begin{table*}[ht]
\caption{Case study of biomedical AI education through PBL in Georgia Tech from 2021 to 2023.}
\resizebox{\textwidth}{!}{
\begin{tabular}{@{}c|cccc|cccc|ccc@{}}
\toprule
Year & \multicolumn{4}{c|}{BHI: Graduate Students}                & \multicolumn{4}{c|}{MIP: Graduate Students}                & \multicolumn{3}{c}{BioStats: Undergraduate Students} \\ \midrule
     & \# students & \# teams & \# publications & Peer Evaluation & \# students & \# teams & \# publications & Peer Evaluation & \# students     & \# teams     & Peer Evaluation     \\ \midrule
2021 & 17          & 5        & 5               & 93.18 \%         & 29          & 8        & 1               & 90.49 \%         & -               & -            & -                   \\
2022 & 22          & 5        & 2               & 89.80 \%         & 28          & 6        & 3               & 90.60 \%         & 46              & 12           & 94.24 \%             \\
2023 & 23          & 7        & 3               & 94.60 \%         & 37          & 10       & 2               & 88.80 \%         & 46              & 9            & 92.62 \%             \\ \bottomrule
\end{tabular}
}
\end{table*}

\subsection{Background: Problem-Based Learning}
PBL is a student-centered educational strategy in which learners utilize authentic problem scenarios (“triggers”) to define learning objectives, investigate knowledge gaps, and collaboratively develop solutions~\cite{mahler2012new, savery2015overview, utley2020interprofessional}. Originating from medical education, PBL emphasizes not only the final solutions but also the learning process itself, mirroring healthcare’s shift toward outcome-based assessment~\cite{hmelo2004problem, wood2003problem}.

PBL typically involves small groups of students working closely together, enhancing communication, teamwork, problem-solving, independent learning, and mutual respect~\cite{wood2003problem}. By engaging with real-world clinical challenges, students directly connect theoretical knowledge to practical clinical applications, preparing them effectively for interdisciplinary biomedical careers~\cite{davis1999amee, barrows1980problem}. Prior work, including Medford et al.’s pilot study at Georgia Tech, demonstrated that vertically integrated PBL teams, combining undergraduate, graduate, and industry participants, successfully addressed complex technical problems (e.g., scoping industry-defined use cases, building reproducible data/ML pipelines, validating against real-world specifications, and packaging results for stakeholder use)~\cite{medford2022online}. However, despite these successes, traditional PBL faces significant challenges such as scalability, particularly in technical fields like biomedical engineering. Typically, such constraints limit PBL activities to theoretical planning or preliminary computational prototypes rather than complete, functional implementations~\cite{dym2005engineering}.

\subsection{Framework: Modularized PBL+AI Model}
To address these limitations, this study introduces a modularized PBL$+$AI framework (Figure~\ref{fig:pbl2}). The design positions GenAI as a knowledge summarization module that accelerates baseline learning and supports code optimization, while preserving the role of PBL in fostering innovation and synthesis. The framework consists of four interdependent modules:

\begin{itemize}
    \item \textbf{Problem Formation in PBL:} This stage begins with defined biomedical problem briefs rather than abstract project selection. Each brief anchors the work in a real-world clinical or research scenario and accompanies curated, deidentified biomedical datasets. These briefs ensure that student inquiry remains problem-based, reproducible, and directly connected to authentic biomedical challenges.
    \item \textbf{Knowledge Inquiry Supported by AI Module:} GenAI tools assist with literature synthesis and coding package recommendations. This module consolidates the state-of-the-art, freeing students to focus on higher-order literature critique and critical thinking. The framework frames AI not as a shortcut to solutions, but as a baseline assembler of existing knowledge.
    \item \textbf{Problem-Solving Module:} Teams advance from knowledge inquiry to AI solution development through implementation. Activities include data quality control, experimental design, model development, and iterative refinement. The curriculum emphasizes the evaluation of robustness and reproducibility to ensure solutions meet the standards of rigor expected in biomedical AI. The protocol permits AI tools to assist with prototyping, but holds teams responsible for team-instrumented debugging and systematic code optimization. Teams must also adhere to disclosure, provenance, and validation requirements.
    \item \textbf{Presentation:} Students consolidate their problem-solving through a thorough justification of AI system output validity, oral communication, peer review evaluation, and manuscript writing. This stage emphasizes higher-order critical thinking, scientific communication, constructive feedback, and reflection on the ethical, clinical, and technical aspects of biomedical AI.
\end{itemize}
This modularized approach reframes AI not as a replacement for the cognitive effort needed for learning but as a scaffold. 
\subsection{Case Study: PBL for Biomedical AI Education}
\subsubsection{Setting and Participants}
The framework was implemented across three academic years (2021–2023) at Georgia Tech in three courses: Biomedical Health Informatics (BHI, graduate level), Medical Image Processing (MIP, graduate level), and Biostatistics (undergraduate level). A total of 248 students participated: 156 graduate students (62 in BHI; 94 in MIP) and 92 undergraduates (Biostatistics). The courses organized students into 62 interdisciplinary teams (41 graduate teams; 21 undergraduate teams) of 3–5 members.

\subsubsection{Team Formation and Support}
Teams were formed using a skill and interest survey to balance prior ML experience, biomedical domain familiarity, and programming proficiency. Instructors and teaching assistants provided weekly lecturing and facilitation to offer formative feedback on problem scoping, methodology selection, evaluation plans, and risk/ethics considerations.

\subsubsection{PBL Workflow, Milestones, and Deliverables}
\begin{itemize}
  \item \textbf{M1: Team Knowledge Inquiry \& Peer Review Feedback (Week 3):}  In this stage, each team selected a specific biomedical AI problem and conducted an in-depth inquiry into relevant literature to understand and critique prior work, state-of-the-art methods, and open research gaps. Teams framed their biomedical problem and technical objectives, subsequently receiving peer feedback through structured reviews. This process helped refine scope, clarify assumptions, and identify potential risks and opportunities.

  \item \textbf{M2: Team Proposal \& Instruction Review Feedback (Week 5):}  Building from M1, each team submitted a formal project proposal which consolidated their problem statement, articulated the data context, proposed detailed AI methodologies and architectures, and outlined anticipated risks, benefits, and ethical/clinical considerations. The proposal also included a preliminary evaluation plan. Instructor feedback at this stage ensured feasibility, alignment with course objectives, and appropriate project scope.

  \item \textbf{M3: Prototype Design, Model Card Draft \& Instruction Review Feedback (Week 7):} Each team built their first functional prototype, provided a rigorous assessment of the results, and drafted a model card documenting data sources, preprocessing steps, modeling choices, assumptions, identified risks, and evaluation criteria. This milestone emphasized critical thinking, transparency, reproducibility, and thoughtful reflection.

  \item \textbf{M4: Final Report \& Demonstration (Week 13):}  Each team produced a comprehensive final package consisting of a written report, reproducible code with documentation, and a short demo presentation. The deliverables included quantitative and qualitative error analyses, ablation studies when feasible, and reproducibility notes highlighting the scientific rigor of the work.
\end{itemize}

Problem briefs addressed authentic biomedical AI tasks across high-impact clinical domains such as diagnostic imaging, physiological sensing, and public health informatics. Examples of projects in the diagnostic imaging domain included leveraging multimodal fusion for Alzheimer’s prediction, implementing surgical video tracking for precision surgery, and performing histopathology segmentation for cancer diagnosis. In the realm of physiological sensing, projects like smartwatch-based COVID-19 detection, eye-tracking for attention deficit hyperactivity disorder assessment, and obstructive sleep apnea detection challenged teams to interpret noisy, longitudinal signals. Furthermore, health informatics projects tackled high-dimensional and unstructured data regarding chronic diseases prevalent in an aging society, with examples ranging from diabetes progression modeling using electronic health records to single-cell Ribonucleic Acid (RNA) sequencing for cancer biomarker discovery and natural language processing of clinical notes. These briefs grounded the milestones in real-world challenges, requiring teams to address domain-specific constraints such as data governance, bias, and clinical validity throughout their inquiries, proposals, prototypes, and final reports.

\subsubsection{AI Tools and Guardrails}
Georgia Tech’s institutional AI-use guidance~\footnote{https://oit.gatech.edu/ai/guidance} permitted the use of GenAI, such as LLM assistants for literature navigation, code optimization strategies, and reflection prompts, and emphasizes privacy, security, and responsible adoption. Each team adhered to the following guardrails:

\begin{itemize}
  \item \textbf{Tool restrictions:} The study prohibited the use of banned software or tools such as \emph{DeepSeek}. Teams used only institutionally approved AI tools that had undergone security and privacy review. 
  \item \textbf{Data protection:} No personally identifiable information (PII), protected health information, or other regulated/confidential Georgia Tech data could be submitted to any AI tool.  
  \item \textbf{Disclosure and validation:}  Teams documented when and how they used AI tools, validated AI outputs for accuracy and bias, and anchored all literature claims in primary sources rather than AI-generated text.
  \item \textbf{Code provenance:} AI-assisted code segments were marked with comments to preserve transparency and reproducibility.  
  \item \textbf{Equity and access:} Approved tools (e.g., Copilot, Zoom AI Companion, Teams Premium) were made available to students to ensure equitable access across teams.  
\end{itemize}
In line with Georgia Tech’s AI strategy, assistance was framed as augmenting PBL stages, such as literature navigation and code optimization strategies, while preserving the integrity of student reasoning and decision-making.

\subsubsection{Measures and Assessments}
The evaluation framework drew on multiple components aligned with the PBL setup: 
\begin{itemize}
  \item \textbf{Team knowledge inquiry (Week 3):} Written reports were evaluated on clarity, completeness of literature critique, and framing of the biomedical and health informatics problems.
  \item \textbf{Proposal (Week 5):} Evaluated for feasibility, methodological alignment, and the articulation of data and ethical considerations.
    \item \textbf{Peer evaluations (midterm and final):} Collected to assess individual contributions, teamwork, and professionalism within each group.
  \item \textbf{Final demo video and oral presentation (Week 13):} Evaluated for technical correctness, clarity of communication, and connection to biomedical decision-making.
  \item \textbf{Final written report (Week 13):}  Evaluated for methodological correctness, result validation, robustness, reproducibility, clarity of communication, and connection to biomedical decision-making.  
\end{itemize}

\subsubsection{Rater Training for Assessment Consistency}
Graduate teaching assistants and the course instructor jointly calibrated grading criteria that aligned project deliverables (knowledge inquiry, proposal, system demo, and final report) with learning outcomes. Periodic discussions among raters monitored assessment consistency, especially for borderline cases.

\subsubsection{Analysis Plan}
Analyses focused on descriptive summaries of course assessments. Peer evaluations were aggregated at the team and individual levels to characterize contributions and communication. The study summarized grades for major deliverables (knowledge inquiry, proposal, demo, final report) across class offerings to assess performance consistency. Narrative reflections and student feedback qualitatively supplemented the quantitative summaries.

\subsubsection{Resources and Infrastructure} 
Compute, software, and storage were provided through institutional resources at Partnership for an Advanced Computing Environment (PACE) at Georgia Tech and industry partners (Microsoft, Hewlett-Packard, and Amazon Web Services). This infrastructure enabled teams to progress beyond theoretical plans to functional prototypes where appropriate. Access policies and cost limits were standardized across offerings.

\subsubsection{Ethical Considerations} 
The coursework design adhered to institutional guidelines for education and research, utilizing deidentified data where applicable. This work provides two significant contributions to the current literature. First, it adapts the PBL methodology to complex biomedical AI education, shifting the focus from theoretical exercises to fully realized computational implementations. This transition aligns the curriculum with modern experiential learning standards. Second, it addresses the inherent scalability limitations of PBL by integrating advanced GenAI technologies, such as LLM-based assistants. These tools facilitate personalized and scalable student learning without substantially increasing faculty workload.

%% file: sec/4-genAI.tex
\section{Results, Challenges and Opportunities}
This section reports the outcome comparison of historical offerings before the adoption of GenAI support in the PBL courses (2016–2017, 2020; Control) versus class offerings after integration (2021–2023; Intervention). Because the 2020–2021 period coincided with the COVID-19 pandemic, which impacted class modality and grading policies, the analysis first examined results including the 2020–2021 data, and then compared results excluding that period. To achieve this, the study prespecified two distributional endpoints: the proportion of A grades (A-rate) and the proportion of low grades (C or D/F), with mean GPA treated as a supportive indicator. When including the COVID-19 year, the Intervention group exhibited a clear rightward shift in the grade distribution, characterized by higher A-rates compared to the Control group (\textit{Control: }39.1\,\% vs. \textit{Intervention:}\ 66.4\,\%; $\Delta=+27.3$ percentage points; Welch $t=4.28$, $p=0.042$; 95\,\% confidence interval (CI) for $\Delta$ $[+2.23,\ +52.43]$) and fewer low grades (\textit{Control: }22.4\,\% vs. \textit{Intervention:}\ 6.1\,\%; $\Delta=-16.3$ points; Welch $t=-5.24$, $p=0.0076$; 95\,\% CI $[-25.18,\ -7.42]$).

Ordinary least squares with heteroskedasticity-consistent errors produced consistent period effects (A-rate: $+27.33$ points, 95\,\% CI $[+12.01,\ +42.66]$; low-grade rate: $-16.30$ points, 95\,\% CI $[-23.76,\ -8.84]$). This indicates a thicker top band and a thinner lower tail during AI-supported PBL offerings. Mean GPA increased from 3.38 to 3.69 ($\Delta=+0.31$; Welch $t=2.30$, $p=0.142$; 95\,\% CI $[-0.245,\ 0.865]$), which is directionally consistent with the distributional shift but underpowered for inference with three-year means per period. Excluding the COVID-19 year revealed that the distributional shift remained: A-rates were higher during the Intervention period (\textit{Control:} 39.1\,\% vs. \textit{Intervention:}\ 67.4\,\%; $\Delta=+28.3$ points; Welch $t=4.36$, $p=0.0367$, 95\,\% CI $[+3.85,\ +52.71]$) and low-grade rates were lower (\textit{Control:} 22.4\,\% vs. \textit{Intervention:}\ 4.25\,\%; $\Delta=-18.12$ points; Welch $t=-6.93$, $p=0.0118$, 95\,\% CI $[-27.66,\ -8.58]$). GPA differences were directionally similar but not significant after the exclusion of the COVID-19 period ($\Delta=+0.325$, $p=0.130$). As a distributional cross-check, a $2\times4$ contingency constructed from year-averaged percentages (A/B/C/D–F) also indicated distinct grade profiles between periods ($\chi^2(3)=22.30$, $p=0.0001$; Cramér’s $V\approx 0.33$), consistent with the other findings.

\begin{figure}[t] 
    \centering
    \begin{subfigure}{}
        \centering
        \includegraphics[width=0.46\linewidth]{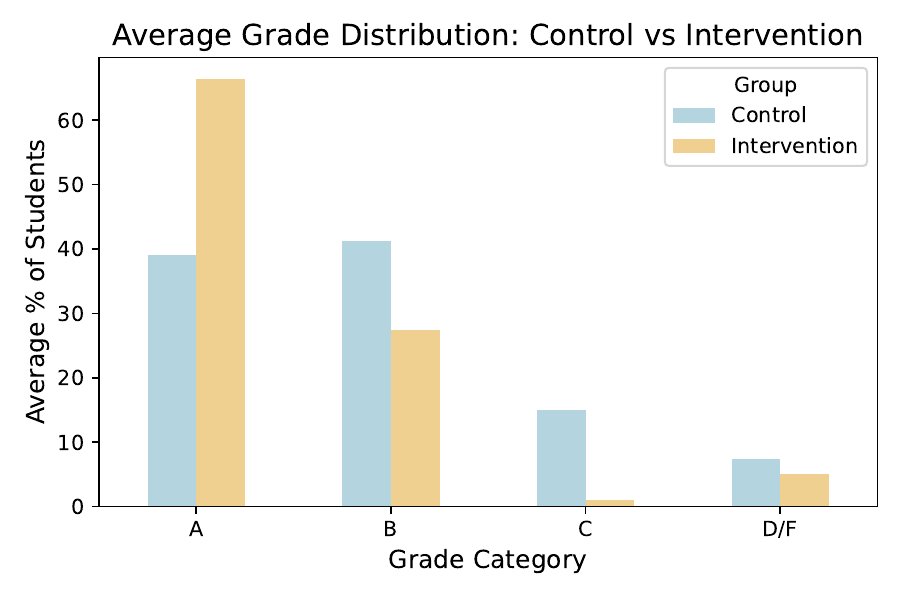}
    \end{subfigure}
    \begin{subfigure}{}
        \centering
        \includegraphics[width=0.46\linewidth]{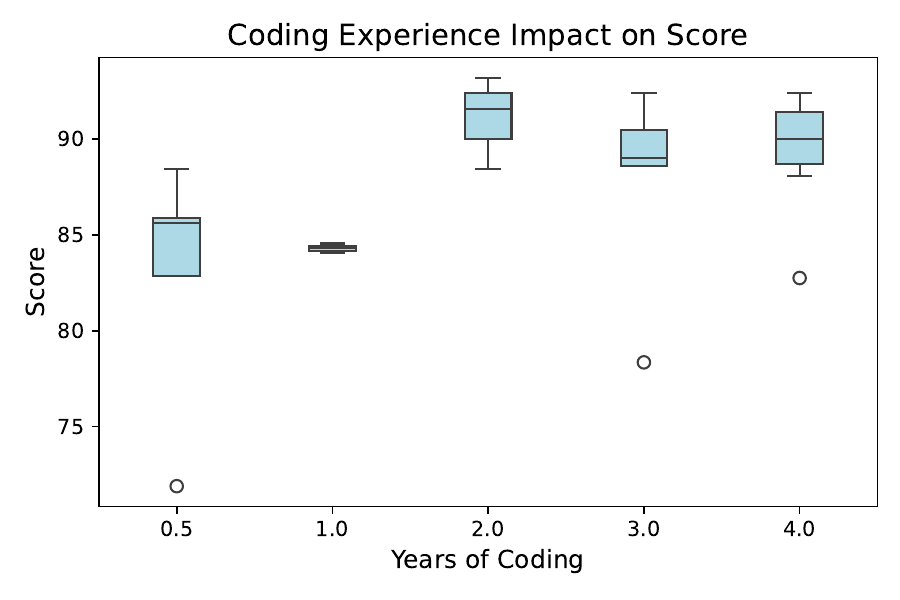}
    \end{subfigure}
    \hfill
    \caption{Effects of GenAI integration and coding experience on performance. (Left) Intervention cohorts with AI access (2021–23) showed significantly improved grade distributions compared to control cohorts without AI (2016–20). (Right) Coding experience's impact on class success.}
    \label{binary_attn_matrices}
\end{figure}

Beyond these cohort effects, Figure~\ref{binary_attn_matrices} visualizes the monotonic association between coding readiness and performance. Students with more years of coding experience tended to perform better, a finding consistent with an exploratory ordinary least squares (OLS) analysis at the student level (slope $\hat\beta=+1.88$ points per year; 95\,\% CI $[+0.30,\ +3.46]$; $p=0.022$). Interpreted conservatively, this pattern suggests that foundational coding fluency is an enabling skill for the effective use of GenAI tooling for ideation and code assistance. This coincides with the observed rightward shift in grades under the AI-supported PBL workflow.

During the Intervention period (2021–2023 in particular), the PBL implementation also yielded sixteen peer-reviewed publications in credible technical conferences by student teams addressing biomedical-AI problems~\cite{tamo2023uncertainty, nnamdi2023model, li2023identification, li2022interpretable, lee2022attention, prabhu2022multi, zamitalo2022development, goel2022identification, gazi2021respiratory, mirzazadeh2021improving, bao2023rare, ding2023personalized, ellis2021novel, ellis2021gradient, ellis2021explainable, usanmaz2021expansion}. Representative examples include a personalized COVID-19 detection framework using wearable signals~\cite{ding2023personalized}, graph-based severity classification from single-cell RNA sequencing~\cite{li2023identification}, and synthetic augmentation for rare transplant-rejection imaging~\cite{bao2023rare}. Consistently, these dissemination outcomes demonstrate that compared to theoretical PBL, implementation-focused PBL aligns with modern experiential learning, improving student engagement and learning outcomes.

Additionally, peer evaluations measured teamwork effectiveness and collaboration, consistently demonstrating high scores across all courses. Specifically, peer evaluations in the BHI course ranged from 89.80\% to 94.60\%, while MIP course scores remained steady between 88.80\% and 90.60\%. The undergraduate Biostatistics course, introduced in 2022, achieved similarly high evaluations (94.24\% in 2022 and 92.62\% in 2023). The observed pattern shift toward higher attainment and fewer low grades under AI-augmented PBL aligns with external evidence that digital or AI-supported PBL improves knowledge and skills over lecture-centric formats and yields gains over traditional PBL. This reinforces randomized and quasi-experimental studies reporting advantages when AI scaffolds are embedded within PBL workflows~\cite{huang2025enhancing,cong2025integrating, de2023using, alreshidi2024effectiveness}. This study’s outcomes are therefore consistent with the direction of effects reported in the literature, though attribution remains limited by the historical, non-randomized design.

\subsection{Discussion: Impact of PBL in Biomedical AI Education}
This study extends prior evidence supporting PBL effectiveness in medical and engineering education~\cite{hmelo2004problem, wood2003problem, freeman2014active, dolmans2005problem} to technical specialty courses. Relative to pre-integration offerings, year-level outcomes under the AI-supported PBL workflow showed a rightward distributional shift (higher A-rates, thinner lower tail) that remained significant even after excluding the pandemic-affected year. The student-level pattern in Figure~\ref{binary_attn_matrices} shows higher performance correlated with greater coding experience, suggesting that coding fluency functions as an enabling condition for the productive use of GenAI tutors and code assistants within PBL. 
\begin{figure}[t]
    \centering
    \includegraphics[width=0.98\linewidth]{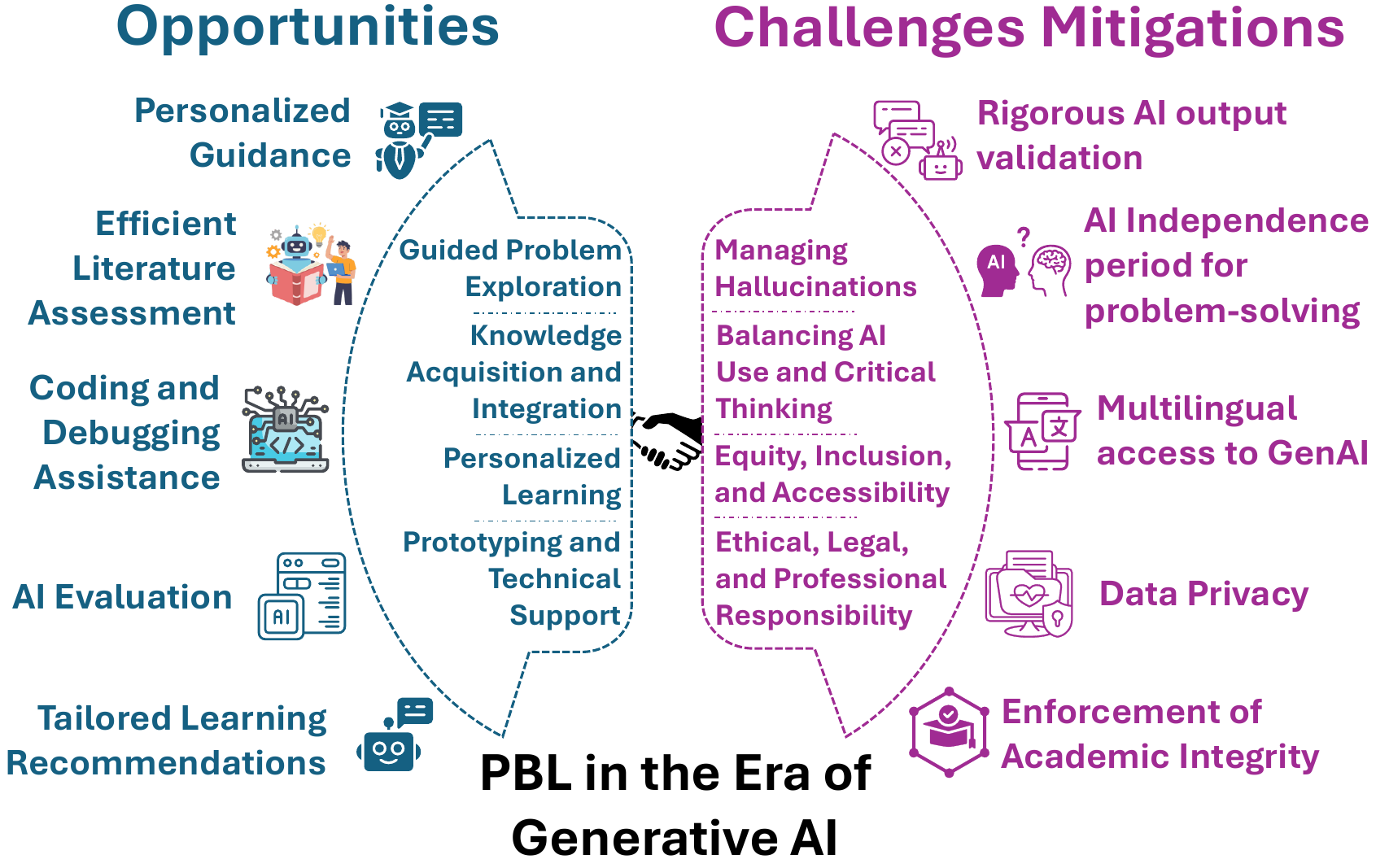}
    \caption{Opportunities, challenges, and mitigations of PBL in the era of GenAI, highlighting pathways for Biomedical AI education}
\label{fig:pbl}
\end{figure}
The modularized PBL+AI architecture offers multiple advantages: (i) by using problem exploration guidance under constrained and disclosed prompting of GenAI, the course offers learner-directed inquiry while reducing the instructor facilitation load; (ii) GenAI offers faster literature synthesis and concept integration, allowing learners to shift time from rote knowledge retrieval to critical thinking and proactive design; (iii) GenAI assists in coding and debugging, enabling learners to iterate from ideation to final manuscript submission within a single academic semester; (iv) the framework integrates formative checks (rubric-aligned auto-feedback, model-card drafts, error-analysis prompts) that increase iteration frequency without expanding grading hours; and (v) the creation of an artifact of problem briefs, milestone rubrics, peer-assessment protocol, model cards, and AI-usage logs, that enables adoption under variable resources.

As a non-randomized study spanning over years, identifying the specific cause-and-effect relationship is challenging. The legacy enrollments constrain enrollment-weighted tests, and thus student-level pre-/post-measures and rater-reliability estimates were not collected in this cycle. In the future, with planning, some refinements, such as enrollment-weighted contingency tests and year-stratified odds ratios can be done if counts are collected. In addition, future work should incorporate student-level ordinal mixed models, pre/post concept and self-efficacy measures, double-rating with quadratic weighted kappa/intraclass correlation coefficient, and standardized AI-usage disclosures with version windows to mitigate tool drift.

\section{Challenges and Opportunities}
Traditional PBL approaches demand considerable faculty resources, domain-specific expertise, and continuous curricular updates to align with rapidly evolving biomedical technologies. With fast-growing AI large foundation models—such as LLMs and vision-language models, these demands can be addressed as summarized in Figure \ref{fig:pbl}. The following sections detail key opportunities alongside challenges and mitigation strategies.

\subsection{Opportunities: Enhancing PBL Scalability with AI}
\noindent $\bullet$ \textbf{Guided Problem Exploration.}
Foundation models can facilitate personalized guidance traditionally provided by faculty. For instance, LLMs can prompt students to critically analyze complex biomedical scenarios and formulate clear learning objectives without prescribing direct solutions~\cite{nazi2024large, waldock2025curriculum}. This approach maintains the learner-directed essence of PBL while significantly reducing faculty facilitation burdens.

\noindent $\bullet$  \textbf{Knowledge Acquisition and Integration.}
Foundation models support the efficient navigation of extensive biomedical literature by summarizing current research, highlighting critical studies, and interpreting complex biomedical concepts and visual information~\cite{bommasani2021opportunities, li2025visual}. This streamlines the self-directed learning phase, enabling students to allocate more time toward critical appraisal and innovative problem-solving.

\noindent $\bullet$  \textbf{Prototype Development and Technical Support.}
GenAI models trained on programming languages can assist students by providing initial computational frameworks, debugging support, and optimization suggestions~\cite{shin2025influence, rouhani2024online}. Consequently, students can focus on higher-order problem-solving, while educators can focus on correct framing of the problem, concept and method selection, and providing constructive feedback beyond routine code implementation checks.

\noindent $\bullet$  \textbf{Automated and Continuous Assessment.}
In addition to assessment by the instruction team, GenAI functions as a learning assistant to provide timely, detailed, and individualized feedback on student progress throughout the PBL cycle. Such automated formative evaluations accelerate learning iterations by timely identification of conceptual misunderstandings and facilitating corrective action~\cite{kim2022teacher, zhan2024integrating}.

\noindent $\bullet$  \textbf{Personalized Learning Pathways.}
GenAI can customize learning experiences by identifying students’ specific learning patterns, strengths, and gaps, and recommending tailored resources and instructional approaches. This personalization addresses challenges arising from varied educational backgrounds common in interdisciplinary biomedical AI courses~\cite{wu2024comprehensive, qin2024tool}.

\subsection{Challenges and Mitigation Strategies}

\noindent $\bullet$  \textbf{Ensuring Accuracy and Managing Hallucinations.}
GenAI models, particularly LLMs, may produce hallucinations and pose risks in biomedical education contexts where accuracy is critical. Mitigation strategies include implementing rigorous validation protocols \cite{mesko2024timely}, integrating retrieval-augmented generation methods that reference verified biomedical literature \cite{thirunavukarasu2024retrieval}, and explicitly training students to evaluate AI outputs critically \cite{masters2022artificial}. These approaches ensure the reliability of AI-generated content in educational settings.

\noindent $\bullet$  \textbf{Balancing AI Use and Critical Thinking Development.}
While GenAI can significantly scale biomedical education, excessive dependence may inhibit students’ deep conceptual understanding and independent critical thinking. Educational strategies should clearly define appropriate use cases for AI assistance~\cite{saqr2023ai}, establish periods of independent problem-solving without AI support, and design assessment methods evaluating AI-generated solutions beyond simple adoption~\cite{han2024artificial}. To foster creativity, instructors should design challenging scenarios to encourage innovative problem-solving and divergent thinking beyond standard AI recommendations \cite{bommasani2021opportunities}.

\noindent $\bullet$  \textbf{Equity, Inclusion, and Accessibility in AI-Enhanced Education.}
GenAI models often display performance biases or reduced effectiveness for non-native English speakers and individuals from diverse cultural backgrounds, potentially reinforcing existing educational inequalities \cite{hovy2023fair}. Institutions must proactively ensure equitable access to AI-enhanced education by developing inclusive prompting strategies, providing multilingual model interfaces~\cite{kaplan2024issues}, and consistently monitoring the performance and effectiveness of AI tools across varied demographic groups \cite{johnson2024integrating}. Such measures are critical for ensuring equitable benefits and reducing disparities in AI education outcomes.

\noindent $\bullet$  \textbf{Ethical, Legal, and Professional Responsibility.}
The educational use of AI technologies presents ethical and legal challenges related to privacy, data security, intellectual property, and academic integrity~\cite{kooli2023biomedical}. Addressing these concerns requires explicit institutional policies governing AI usage in educational settings~\cite{wcet2023developing}, robust frameworks for data management, and clear delineation of responsibilities regarding AI-generated content. Transparency and clear communication regarding the role of AI in educational processes are essential to maintaining academic integrity and upholding professional and ethical standards in biomedical education~\cite{cordero2024advancement}.

By carefully leveraging these AI technologies and proactively addressing their associated challenges, BME programs can significantly enhance the scalability and effectiveness of PBL, preparing a diverse student body for healthcare in the era of AI.  Table~\ref{tab:dataset} provides a summary of publicly available datasets and benchmarks commonly used for evaluating biomedical educational applications of LLMs.

\begin{table}
\centering
\caption{Summary of existing datasets and benchmarks for evaluating LLMs on biomedical education applications.}
\label{tab:dataset}
\resizebox{0.49\textwidth}{!}{
\begin{tabular}{cccccc}
\hline
Dataset\&Benchmark & Subject & Modality & Amount \\
\hline
MedQA~\cite{jin2021disease}              & medicine            & text          & 40 K   \\
MedMCQA~\cite{pal2022medmcqa}            & medicine            & text          & 200 K  \\
MMLU (med)~\cite{hendrycks2020measuring} & medicine         & text          & 272 K  \\
PubMedQA~\cite{jin2019pubmedqa}           & biomedicine         & text          & 1 K    \\
MedQA-USMLE~\cite{jin2021disease}        & medicine            & text          & 12.7 K \\
BioASQ~\cite{tsatsaronis2015overview}             & biomedicine         & text          & 3 K    \\
HeadQA~\cite{vilares2019head}             & healthcare          & text          & 2.7 K  \\
MedDialog~\cite{he2004meddialog}          & medicine            & text          & 3.6 M  \\
PathVQA~\cite{he2020pathvqa}            & pathology           & text \& image & 32 K   \\
VQA-RAD~\cite{lau2018dataset}            & radiology           & text \& image & 3.5 K  \\
SLAKE~\cite{liu2021slake}              & medicine            & text \& image & 14 K   \\
\hline
\end{tabular}}
\end{table}

%% file: sec/6-conclusion.tex
\section{Conclusion}
This study operationalized a PBL framework tailored to technical specialty courses and examined outcomes across historical cohorts before and after the integration of GenAI support. Post-integration offerings showed a rightward shift in grade distributions (higher proportions of A grades, fewer low grades) that remained robust after excluding the pandemic-affected year. Exploratory student-level signals suggest that compared to theoretical PBL, implementation-focused PBL aligns more closely with modern experiential learning, resulting in improved student engagement and learning outcomes. The primary output of this work is a modularized, guard-railed PBL+AI architecture accompanied by a portable implementation package (problem briefs, milestone rubrics, peer-assessment protocols, model cards, and AI-usage logs). The Discussion details limitations inherent to the historical design and incomplete legacy enrollments. Future iterations will incorporate enrollment-weighted tests, student-level mixed models, and pre-/post-concept and reliability measures. Overall, AI-supported PBL appears to elevate attainment while preserving core PBL principles. This approach aligns with modern experiential learning standards and better prepares biomedical engineering students to engage responsibly with healthcare in the AI era.